\documentclass[doublecol]{epl2}
\usepackage{epsfig}
\usepackage{graphicx}
\usepackage{bm}
\newcommand{\simleq}{\mbox{\raisebox{-1.0ex}{$\stackrel{<}{\sim}$}}}

\newcommand{\Dr}{\Delta_{r}}
\newcommand{\Dren}{\Delta_{ren}}

\newcommand{\alf}{\alpha_b}

\newcommand{\wco}{\omega_{co}}
\newcommand{\wo}{\omega_{0}}

\title{Dissipation due to two-level systems in nano-mechanical devices}

\author{C. Seoanez\inst{1} \and F. Guinea\inst{1} \and A.~H. Castro Neto\inst{2,3}}

\institute{
  \inst{1} Instituto de  Ciencia de Materiales de Madrid, CSIC - Cantoblanco
  E28049 Madrid, Spain\\
  \inst{2} Department of Physics, Harvard University - Cambridge, MA 02138,USA\\
  \inst{3} Department of Physics, Boston University - 590 Commonwealth
  Avenue, Boston, MA 02215, USA }


\abstract{We analyze the dissipation of the vibrations of
nano-mechanical devices. We show that the coupling between flexural
modes and two-level systems leads to sub-ohmic dissipation. The
inverse quality factor of the flexural modes of low frequencies depends on
temperature as $Q^{-1} ( T ) \approx Q_0 + C T^{1/3}$, providing a
quantitative description of the experimental data.}

\begin{document}

\maketitle

\section{Introduction}

Nano-electro-mechanical devices \cite{C02,B04} (NEMS) are systems
with great potential for applied physics and engineering because of
their extreme sensitivity, as probes, to their environment
\cite{AGHR91,Setal97,Ietal00,LD03,EYR04,BBMC04,DKEM06}. Furthermore,
because of their small size and large surface to volume ratio, these
systems are in the crossover region between classical and quantum
behavior, and hence of great theoretical interest. Thus, the study
of the sources of noise and dissipation in these systems has
attracted a great deal of attention
\cite{CR99,Yetal00,Eetal00,CR02,YOE02,Metal02,AM03,Hetal03,ZGSBM05}.
One of the common realizations of nano-mechanical resonators is a
rigid beam of nanoscopic dimensions which vibrates at GHz
frequencies \cite{C00,GZBM05}. The damping of these oscillations has
been a subject of intense investigation
\cite{CR99,CR02,Metal02,AM03}, as it sets a limit to their possible
applications.

The damping of a low frequency oscillation in a NEMS comes from the
coupling of this mode to other low energy degrees of freedom.
Experiments suggest that surfaces of nano-mechanical resonators
resemble amorphous bulk systems \cite{LTWP99}, with a high density
of defects, playing a major role as a source of dissipation
\cite{YOE00}. In amorphous solids, disorder and impurities lead to
the existence of anharmonic excitations, which can be modeled as a
degree of freedom tunneling between two potential wells \cite{RH05}.
A simpler two-level system (TLS) description arises at low
temperatures, when only the two lowest eigenstates have to be
considered. Some properties of the distribution of TLSs in terms of
their parameters (bias, $\Delta_{0}^{z}$, and tunneling rate,
$\Delta_{0}^{x}$) can be inferred from experiments \cite{E98}, and
they are considered to be the main source of damping of acoustical
modes in disordered insulating solids \cite{AHV72,P72,P87,E98}.

\begin{figure}[t]
  \begin{center}
      \includegraphics[width=4cm,angle=-90]{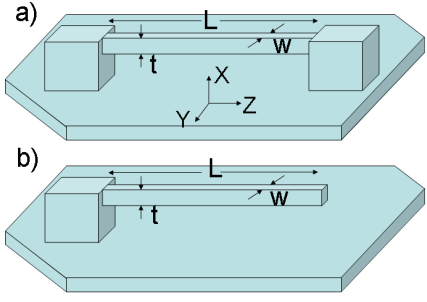}
\end{center}
   \caption{Sketch of the NEMS: a) Doubly clamped beam, suspended
      by both ends; b) Cantilever, with one free end. The
      device is characterized by its  width  ($w$), thickness
      ($t$), and length, ($L$), where $w \sim t \ll L$.}
    \label{sketch_beam}
\end{figure}

In this work, we study  mostly a rigid beam geometry, sketched in
fig.~[\ref{sketch_beam}], and analyze the dissipation processes for
low energy flexural modes due to the presence of effective TLSs at
its surface. The generalization to dissipation of torsional modes is
straightforward and will be given elsewhere. Once a given mode is
externally excited, the TLSs living at the surface of the beam,
which are coupled to this mode, will absorb part of its energy. But
as the TLSs are also coupled to the rest of vibrational modes of the
beam, they will release most of this energy to them. Thus the TLSs
give rise to an indirect coupling between the externally excited
mode and the rest of modes. For the experimentally relevant case of
low amplitudes of vibration this coupling prevails over the usual
anharmonic coupling.

This energy flow process will be described in two stages:

i) From the point of view of a given TLS, its coupling to the
vibrational modes of the beam, which can be seen as an external
bath, alters its dynamics and enables it to absorb and emit energy
in a broad range of frequencies. In particular, the presence of
flexural modes leads to the possibility of qualitative changes in
the dynamics of the TLSs, as the former constitute a {\em sub-ohmic}
environment \cite{Letal87,W99} for the TLSs. This is a consequence
of the quadratic dispersion relation characteristic of flexural
modes, which results in an enhancement of the density of low
frequency modes.

ii) Coming back to the externally excited vibrational mode whose
damping we want to compute, the TLSs, dressed by all the vibrational
modes of the structure, constitute the dissipative environment for
the mode. This involves TLSs that are nearly resonant with the mode
under consideration but also off resonance TLSs since the strong
phonon/TLS coupling provides each TLS spectral function with tails
far from resonances. In many NEMS experiments, the flexural modes
studied are highly excited, either because of an external driving
mechanism, or because the temperature is much higher than the
frequency of the mode. As a given TLS can absorb and emit over a
broad range of energies, due to the incoherent tails in its spectrum
(see below), these processes allow for the transfer of energy from a
highly excited low frequency flexural mode to other modes with
higher frequencies.

Dissipative mechanisms other than TLSs, which have been extensively
studied elsewhere \cite{JI68,CL01,PJ04}, are not considered here.

\section{TLSs coupled to low dimensional vibrations}

The TLSs are mainly coupled to the strain induced by phonons. It is
assumed that the main effect of the strain is to modify the energy
splitting between the TLSs energy levels\cite{A86}. The Hamiltonian
of a given TLS is characterized by a bias, $\Delta_{0}^{z}$, and the
tunneling rate, $\Delta_{0}^{x}$ (we use units such that
$\hbar=1=k_B$, and omit the part of the Hamiltonian describing free
vibrations):
\begin{equation}\label{Hamiltonian}
    \textsl{H}=\Delta_{0}^{x}\sigma_x+\Delta_{0}^{z}\sigma_z+\sigma_z\textsl{F}(\partial_i
    u_j)
\end{equation}
where $ \partial_i u_j $ is a component of the deformation gradient
matrix, and $\textsl{F}$ is an arbitrary function. Changing basis to
the energy eigenstates of the TLS, eq.(\ref{Hamiltonian}) becomes
$\textsl{H}= \Delta_{0}\sigma_z +
    [(\Delta_{0}^{x}/\Delta_{0})\sigma_x+(\Delta_{0}^{z}/\Delta_{0})\sigma_z]\textsl{F}(\partial_i u_j)$.
 $\Delta_0=\sqrt{(\Delta_{0}^{x})^2+(\Delta_{0}^{z})^2}$ is the splitting of the TLS.
Dissipation is dominated by slightly biased TLSs for whom
$\Delta_{0}^{z}\ll\Delta_{0}$, so the last term can be ignored. A
further expansion of \textsl{F} to lowest order in the displacement,
together with a $\pi/4$ rotation of the eigenbasis, leads to: $
   \textsl{H}= \Delta_{0}\sigma_x +
    \gamma (\Delta_{0}^{x}/\Delta_{0})\sigma_z \partial_i u_j
$ , where $\gamma$ is the coupling constant (with dimensions of
energy).

The main interaction between phonons and TLSs is due to the coupling
to the operator $\sigma_z$ of each TLS. Hence, the absorption
properties of each TLS can be characterized by the spectral
function:
\begin{equation}
A ( \omega ) \equiv \sum_n \left| \left\langle 0 \left|
\sigma_z \right| n
  \right\rangle \right|^2 \delta ( \omega - \omega_n + \omega_0 )
\label{spectral_TLS}
\end{equation}
where $| n \rangle$ is an excited state of the total system TLS
plus vibrations. The linearization of the coupling implies that the
interaction between a given TLS and the vibrations can be written as
$\textsl{H}_{int} \equiv \sigma_z \sum_k \lambda_k \left( b_k^\dag +
b_k \right)$.

We also define a spectral function that determines the
damping induced by phonons on the TLS: $ J ( \omega ) \equiv \sum_k
\left| \lambda_k \right|^2 \delta (
\omega - \omega_k ) $ where $\omega_k$ is the energy of mode $k$.\\
The vibrational modes of a beam with fixed ends have a discrete
spectrum, but we will approximate them by a continuous distribution.
This approximation will hold as long as many vibrational modes
become thermally populated, $kT\gg\hbar\omega_{fund}$, where
$\omega_{fund}$ is the frequency of the lowest mode. The condition
is fulfilled in current experimental setups.

\subsection{Acoustic modes of a nanoscopic beam}

Using continuum elasticity theory \cite{LL59}, a one-dimensional
(1D) rod has compression and twisting modes, with a linear relation
between frequency and momentum, and bending, or flexural, modes,
where the frequency depends quadratically on momentum. We will
consider a rod of length $L$, width $w$ and thickness $t$, see
Fig.[\ref{sketch_beam}]. We describe next the spectral function which
describes how the modes of the rod absorb energy in different energy ranges.

The compression and twisting modes lead to an ohmic spectral
function for $\omega \ll 2\pi c/R$ ($R$ being a typical transversal
dimension of the rod and $c$ the sound velocity), when the rod is
effectively 1D. In terms of the Young modulus of the material, $E$,
and the mass density, $\rho$, we get: $J_{{\rm comp}}( \omega ) =
\alpha_c |\omega|$, where,
\begin{equation}
    \alpha_c= (\gamma \Delta_{0}^{x}/\Delta_{0})^2
(2 \pi^2 \rho t w )^{-1} (E/\rho)^{-3/2} \, .
\end{equation}
The twisting modes are defined by the torsional rigidity, $C = \mu
t^3w/3$ ($\mu$ is a Lande coefficient), and $I=\int dS x^2= t^3w/12$
(where $S$ is the cross-section). The corresponding spectral
function is given by: $J_{{\rm torsion}}(\omega) = \alpha_t
|\omega|$, where
\begin{eqnarray}
\alpha_t = C (\gamma \Delta_{0}^{x}/\Delta_{0})^2
 (8\pi^2 \mu t w  \rho I)^{-1} (\rho I/C)^{3/2} \, .
\end{eqnarray}

The analysis of the flexural (bending) modes differs substantially
from the other ones, because they correspond to two fields
$\Phi_j(z,\omega)$ ($j=x,y$) that satisfy \cite{LL59}: $
   E I_{j} \partial_z^4 \Phi_j = \rho \, t \,  w \, \omega_j^2 \, \Phi_j \, ,
$ where, for the system considered here, $I_{j} = t^3 w/12$. The
normal modes have a quadratic dispersion
  $\omega_{j}(k) = \sqrt{E I_j/(\rho tw)} \, k^2$.
Their corresponding  spectral function is sub-ohmic \cite{Letal87},
$J_{{\rm flex}}(\omega) = \alpha_b\sqrt{\wco} \sqrt{\omega}$, with,
\begin{eqnarray}\label{Jsubohmic}
 \alpha_b \sqrt{\wco} =
 0.3\frac{\gamma^2}{t^{3/2}w}\frac{(1+\nu)(1-2\nu)}{E(3-5\nu)}
   \Bigl(\frac{\rho}{E}\Bigr)^{1/4} \, ,
\end{eqnarray}
where $\nu$ is Poisson's ratio and $\wco\simeq \sqrt{EI_y/(\rho
tw)}(2\pi/t)^2$ is the high energy cut-off of the bending modes.
\begin{figure}[t]
  \begin{center}
\includegraphics[width=4cm]{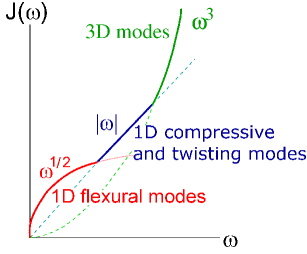}
\end{center}
    \caption{Sketch of the contributions to the spectral function which
      determines the dynamics of the TLS.}
    \label{spectral_function}
\end{figure}
Collecting the previous results, we find the spectral function
$J(\omega)$ plotted in Fig.[\ref{spectral_function}].

\section{Dynamics of the TLSs}

\begin{figure}[t]
  \begin{center}
\includegraphics[width=4 cm,angle=-90]{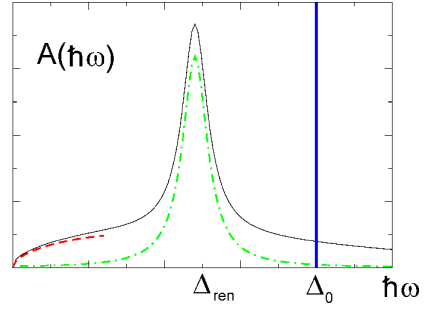}
\end{center}
    \caption{Sketch of the spectral function of a TLS coupled to a sub-ohmic
      bath. Dashed line shows the off-resonant contribution. Dot-dashed line
      shows the main broadened peak. For comparison, the thick vertical line
      shows the spectral function of a non-interacting TLS.}
    \label{absorption_TLS}
\end{figure}

We are interested in the TLSs that affect most the low-energy
flexural vibrations. Hence, we focus on the TLSs whose tunneling
amplitudes lie in the region where the damping is sub-ohmic. We
define $\Dr$ as the tunneling amplitude of a given TLS, including
the renormalization due to the high energy acoustic modes. We
assume, as in the case of glasses, that the distribution of these
TLSs is given by \cite{AHV72,E98} $g(\Delta_r,\Delta_z)=P/\Delta_r$.
The sub-ohmic coupling, eq.(\ref{Jsubohmic}), leads to a
renormalization of $\Dr$:
\begin{equation}
\Dren = \Dr \exp\{- \alpha_b \sqrt{\wco}
\smallint^{\omega_{co}}_{\Dren} d \omega J (\omega)/\omega^2\}
\end{equation}
This equation has no solutions other than $\Dren = 0$ if
 $\Dr \ll \alpha_b^2 \wco$, so that the tunneling amplitude of the low energy TLSs
is strongly suppressed\cite{SD85,Letal87,W99,KM96}. The remaining
TLSs experience a shift and a broadening of the spectral function
function $A ( \omega )$, defined in eq.(\ref{spectral_TLS}). In
addition, $A ( \omega )$ acquires
 a low energy tail, which,
at zero
 temperature, is\cite{G85}:
\begin{equation}
A ( \omega ) \propto
\alpha_b \frac{\sqrt{\wco \omega}}{ \Dren^2} \, \, \, \, \, \, \, \, \omega
\ll \Dren
\label{off_res_TLS}
\end{equation}
There is also a high energy part, $A ( \omega ) \propto \alpha_b
\sqrt{\wco} \Dren^2 \omega^{-7/2}$, for  $\omega \gg \Dren$. The
main features of $A ( \omega )$ are shown in
fig.~[\ref{absorption_TLS}]. Finally, we obtain the width of
 the resonant peak, $\Gamma ( \Dren )$, using Fermi's golden rule,
 $\Gamma(\Dren)= 16 \alpha_b\sqrt{\wco} \, \sqrt{\Dren}$.
This description is valid for wavelengths such that, $1/L \ll k \ll
1/{\rm max}(w,t)$.

The value of $\Delta_z$ is not renormalized by the phonons, so that
the TLS cannot exchange energy with the environment at frequencies
lower that $\Delta_z$. In the following, we will consider only TLSs
with $\Delta_z \simleq \Dren$.

The total absorption rate by the ensemble of dressed TLSs present in
the beam, $A_{tot} ( \omega )$, is obtained summing over
$g(\Delta_r,\Delta_z)$ the values of $A ( \omega )$,
eq.(\ref{spectral_TLS}), of each dressed TLS. In amorphous
insulators, like amorphous silica, there are TLSs with $\Delta_r$ up
to about $\omega^*=$5 K \cite{E98}. The upper cut-off of the
distribution is usually larger than the frequencies of the flexural
modes of interest, $\omega_{co}<\omega*$. Integrating over
$\Delta_z$ and $\Dren$, we find that the density of TLSs per unit
volume and unit energy available for direct (resonant) excitation
processes is given by $P$, in agreement with the known result that
TLSs in amorphous systems give rise to a finite density of states at
low energies\cite{AHV72}. In addition, we find a contribution coming
from the non-resonant part of the spectral function of each TLS,
$A_{tot}^{off-res}(\omega) \approx 2 P \alpha_b \sqrt{\wco /
\omega}$. The divergence as $\omega \rightarrow 0$ arises from the
$\Dren^{-2}$ dependence of $A ( \omega )$, eq.(\ref{off_res_TLS}).
\section{Dissipation due to the TLSs}

We assume that the externally excited flexural mode of interest,
$(k_0,\wo)$, is linearly coupled, with the coupling constant shown
in the derivation of the hamiltonian $ \textsl{H}$, $\gamma
(\Delta_{0}^{x}/\Delta_{0})$, to a continuum of excitations whose
spectral strength, $A_{tot} ( \omega )$,  is given by the summation
over all TLSs of the function $A ( \omega )$ calculated for each
one. The ratio $(\Delta_{0}^{x}/\Delta_{0})$ can be approximated as
1, due to the negligible role played by strongly biased TLSs.

The transition rate
of the mode $k_0$ occupied by $n$ phonons to the mode with $n-1$ is
calculated from Fermi's golden rule, and the energy loss per cycle
and unit volume $\Delta E$ of the mode will correspond to this
transition rate multiplied by the energy of a phonon $\hbar\wo$ and
the period $2\pi /\wo$:
\begin{equation}\label{FGRrate}
    \Delta E\simeq \frac{2\pi}{\wo}\times\hbar\wo\times\frac{2\pi}{\hbar}n\left(\gamma
\frac{k_{0}^{2}}{\sqrt{\wo}}\right)^2 A_{tot}(\wo)
\end{equation}
where $A_{tot}(\omega) = P + P \alpha_b \sqrt{\wco / \omega}$ is the
sum of the resonant and non resonant contributions arising from
integrating over the distribution of TLSs, as discussed in the
preceding paragraph.

The inverse quality factor $Q^{-1}(\omega_0)$ is given by
$Q^{-1}(\omega_0)=\Delta E / 2\pi E_0$, where $E_0$ is the energy
stored in the mode per unit volume, $E_0\simeq n\hbar\wo / twL$,
leading to the following expression at zero temperature:
\begin{equation}\label{Mainresult}
Q^{-1}(\omega_0) \simeq 10 t^{3/2} w \left(\frac{E}{\rho}\right)^{1/4}
\alpha_b \sqrt{\wco} A_{tot}(\omega_0)
\end{equation}

\begin{figure}[t]
\begin{center}
\includegraphics*[width=.99\columnwidth]{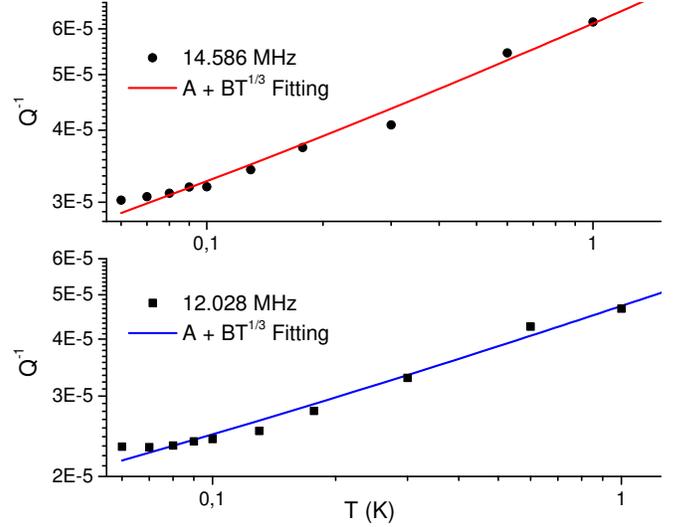}
\end{center}
    \caption{Fittings of the result of eq.(\ref{totaloffresonance})
to experimental data  \cite{ZGSBM05}.}
    \label{Fittings}
\end{figure}

Experiments are done at finite temperatures, and, sometimes, in
systems where the oscillator is driven strongly out of equilibrium.
We take these effects into account by calculating $A_{tot} ( \omega
, T )$ in the presence of a finite distribution of excited
vibrations, which can include a non-thermal contribution. If we only
keep one phonon processes in the calculation of $A_{tot} ( \omega ,
T )$, as in the previous discussion, the subtraction of absorption
and emission processes cancel the temperature dependence on the
number of phonons. The only temperature dependence is due to
saturation effects in the absorbtion properties of the TLSs when
their environment contains many quanta of energy $\Dren$. The final
expression for the inverse quality factor at finite temperatures is:
\begin{eqnarray}\label{QT}
    Q^{-1}(\omega_0 ,T) &\simeq & 10 t^{3/2} w
    \left(\frac{E}{\rho}\right)^{1/4} \alpha_b \sqrt{\wco}  \times
    \nonumber \\ &\times &
\left[ P \tanh\left(\frac{\hbar\omega_0}{k_BT}\right) + P \alpha_b
    \sqrt{\frac{\wco}{\omega_0}} \right]
\end{eqnarray}
Until now prevalence of one-phonon processes in the interaction
among TLSs and vibrational modes has been assumed, but at
temperatures much higher than the frequencies of the relevant
phonons, multi-phonon processes need to be taken into account. We
include this effect assuming overdamped dynamics for the TLSs, so
that $A(\omega,\Dren,T)= \tau(\Dren,T)/(1+[\omega\tau(\Dren,T)]^2)$,
where $\tau(\Dren,T)=\Gamma^{-1}(\Dren,T)$, and making use of the
relation between $Q^{-1}(\omega,T)$ and $A(\omega,\Dren,T)$ usually
found in the context of the standard tunneling model approach to
disordered bulk systems \cite{AHV72,P72,J72,P87,E98}, which in our
case translates into:
\begin{eqnarray}\label{QEsq}
    Q^{-1}(\omega,T)&=& P\gamma^2/(E T)
    \smallint_{0}^{\epsilon_{max}}d\epsilon
\smallint_{u_{min}}^{1} du \sqrt{1-u^2}/u \times \nonumber \\
&\times & \frac{\cosh^{-2}\left(\frac{\epsilon}{2T}\right) \,
    \omega\tau}{1+(\omega\tau)^2}
\end{eqnarray}
Here $\epsilon=\sqrt{\Dren^2+\Delta_{z}^{2}}$ and
$u=\Dren/\epsilon$, with $u_{min}$ fixed by the time needed to
obtain a spectrum around the resonance frequency of the excited mode
and $\epsilon_{max}$ estimated to be at least of the order of 5 K
\cite{E98}. The limits of integration must be such that only the
overdamped TLSs are included. A given TLS is overdamped when $\Dren
\leq \Gamma(\Dren,T)\rightarrow \Dren \leq [30 \alf\sqrt{ \wco}
T]^{2/3}$. Hence, we obtain for the contribution to the inverse
quality factor from overdamped TLSs:
\begin{equation}\label{QEsq2}
 Q^{-1}(\omega_0,T) \approx
 \frac{400 P \gamma^2 (\alf\sqrt{ \wco})^{4/3} \, T^{1/3} }{E\omega_0 } \, .
\end{equation}
Therefore, in a range of energies $\omega_{fund} \leq \omega_0 \ll
\wco$, the attenuation coming from these TLSs shows a $Q^{-1}(T)\sim
T^{1/3}$ dependence, in qualitative agreement with the experiment
\cite{ZGSBM05} on Si nanobridges. The total inverse quality factor
is the sum of eq.(\ref{QEsq2}) plus the temperature independent
contribution arising from off-resonant processes induced by
underdamped TLSs, eq.(\ref{QT}). This last equation must in
principle be corrected by taking into account the decrease in the
number of underdamped TLSs as the temperature is raised, but it is a
weak effect and will be neglected, leading to the final expression
for the total attenuation of a flexural mode:
\begin{equation}\label{totaloffresonance}
    Q^{-1}(\omega_0,T)\sim Q_0^{-1}(\omega_0)+C(\omega_0)  T^{1/3} \, ,
\end{equation}
where,
\begin{eqnarray}
  \nonumber Q_0^{-1}(\omega_0) &\simeq& \frac{3 P\gamma^4\rho^{1/4}}{t^{3/2} w
E^{9/4}} \Bigl[\frac{(1+\nu)(1-2\nu)}{3-5\nu}\Bigr]^2 \omega^{-1/2}_0 \\
  C(\omega_0) &\simeq& \frac{150 P \gamma^4\rho^{1/3}}{t^2w^{4/3} E^4 \omega_0}
\Bigl[\frac{(1+\nu)(1-2\nu)}{3-5\nu}\Bigr]^{4/3}
\end{eqnarray}
As only part of the beam is amorphous, $P$ is to be replaced by
$P\cdot V_{{\rm amorphous}} / V_{{\rm total}}$. To describe the
results in \cite{ZGSBM05}, as shown in fig.~[\ref{Fittings}], we
assumed $P\cdot V_{{\rm amorphous}} / V_{{\rm
total}}\sim10^{44}$J$^{-1}$m$^{-3}$, compatible with $P$ values
reported in amorphous glasses \cite{E98}, and $0.1 \simleq V_{{\rm
amorphous}} / V_{{\rm total}} \simleq 1$. The slope of the $T^{1/3}$
contribution gives the value of $\gamma$, used as fitting parameter.
We obtain $\gamma\sim 5-10$ eV, which is a reasonable value
\cite{P88,K89}. There are two limitations on the range of
applicability of our results. The increasing role of interactions
between TLSs as T is lowered \cite{ERK04}, and possible cooperative
effects when the mode is strongly driven \cite{AM03} cause
deviations, which are manifested in the saturation observed in
fig.(4), not explained by our fit, which predicts
$Q_{0}^{-1}\sim5\cdot10^{-7}$. On the high-temperature side, a point
is reached when the rate $\Gamma^{-1}(\Dren,T)$ changes to an
Arrhenius-like behavior \cite{E98}.

\section{Conclusions}

We have studied the damping of mechanical oscillations in nanoscopic
devices due to their interaction with TLSs. This coupling is the
main mechanism of relaxation of phonons in disordered insulators. We
have analyzed the changes induced in the spectrum and distribution
of TLSs due to their interaction with the low energy oscillations of
nano-mechanical devices. Flexural modes, with a high density of
states at low energies, lead to sub-ohmic damping, which can modify
significantly the distribution of TLSs. The problem of a TLS
interacting with a sub-ohmic environment is interesting in its own
right \cite{Letal87,KM96,W99,S03,VTB05,GW88,SG06,ChT06,K04,IN92},
and the systems studied here provide a physical realization.

We obtain a temperature independent contribution to the inverse
quality factor, $Q^{-1}$ of a flexural mode, which arises from
resonant excitations of TLSs, and off-resonant processes involving
underdamped TLSs. We find, in addition, a contribution which
increases as $T^{1/3}$, arising from overdamped TLSs. The
off-resonant contributions imply that the externally excited
vibration loses its energy to TLSs which, in turn, decay into other
acoustic modes. Hence, off-resonant contributions can only be
present if the number of thermally excited modes is large, a
situation fulfilled in most present experiments.

We have made numerical estimates for the expected dissipation for a
few representative devices.  We have assumed that a fraction of the
device shows amorphous features, and contains a distribution of TLSs
similar to that found in amorphous insulators. The main
uncertainties in our calculation are due to the lack of information
on the TLSs distribution, the coupling strength, and the fraction of
the total volume of the device that they occupy. Decreasing volume
and number of modes may lead as well to fluctuations around our
predictions, which use continuum distributions.

\acknowledgments We acknowledge many helpful discussions with P.
Mohanty. C. S. and F. G. acknowledge funding from MEC (Spain)
through grant FIS2005-05478-C02-01 and the Comunidad de Madrid,
through the program CITECNOMIK, CM2006-S-0505-ESP-0337. A.H.C.N. is
supported through NSF grant DMR-0343790.

\end{document}